\documentclass[twocolumn,pre]{revtex4}
\usepackage{graphics}
\usepackage{graphicx}
\usepackage{amsfonts}
\usepackage{textcomp}
\usepackage{amssymb}
\usepackage{mathrsfs}
\usepackage{amsmath}
\usepackage{color}
\usepackage{ulem}
\usepackage{float}

\begin{document}
\title{Hyperuniform Density Distributions of Brownian Particles via Designer External Potentials}

\author{Yang Jiao}
\email[correspondence sent to: ]{yang.jiao.2@asu.edu}
\affiliation{Materials Science and Engineering, Arizona State
University, Tempe, AZ 85287} \affiliation{Department of Physics,
Arizona State University, Tempe, AZ 85287}



\begin{abstract}

Disordered hyperuniformity (DHU) is a recently discovered novel state of many-body systems that is characterized by vanishing normalized infinite-wavelength density fluctuations similar to a perfect crystal, yet possesses an amorphous structure like a liquid or glass. Due to their unique structural characteristics, DHU materials are typically endowed with unusual physical properties, such as large isotropic photonic band gaps, optimal transport properties and superior mechanical properties, enabling a wide spectrum of novel applications. Here we investigate equilibrium DHU states of Brownian particles induced by external potentials. In particular, we analytically derive sufficient conditions on the external potentials in order to achieve distinct classes of DHU density distributions of Brownian particles in thermal equilibrium, based on the stationary-state solutions of the corresponding Smoluchowski equation. We show for a wide spectrum of tight-binding potentials, the desirable DHU states of Brownian particles can be controlled and achieved by imposing proper hyperuniformity conditions on the potentials. Moreover, we find that thermal motions in these systems tend to enhance hyperuniformity. We also analyze the evolution dynamics of an initial density distribution (hyperuniform or non-hyperuniform) to the desirable equilibrium DHU state determined by the prescribed external potentials, which is shown to be coupled with the full spectra of the force fields associated with the imposed potentials. We find that although the transient density distribution can rapidly develop local patterns reminiscent of those in the equilibrium distribution, which is governed by the fast dynamics induced by the external potential, the overall distribution is still modulated by the initial density fluctuations which are relaxed through slow diffusive dynamics. Our study has implications for the fabrication of designer DHU materials.


\end{abstract}
\maketitle

\section{Introduction}



Disorder hyperuniformity (DHU) is a recently discovered novel
state of many-body systems \cite{To03, To18a}, possessing a
hidden order in between that of a perfect crystal and a totally
disordered system (e.g., an ideal gas). DHU systems are statistically isotropic and possess no Bragg peaks, yet they suppress
large-scale density fluctuations like crystals \cite{To03},
which is manifested as the
vanishing static structure factor in the infinite-wavelength (or
zero-wavenumber) limit, i.e., $\lim_{k\rightarrow 0}S(k) = 0$,
where $k$ is the wavenumber. DHU is equivalently
characterized by a local number variance $\sigma_N^2(R)$ associated
with a spherical window of radius $R$ that grows more slowly than
the window volume (e.g., with scaling $R^d$ in $d$-dimensional
Euclidean space) in the large-$R$ limit \cite{To03, To18a}. The small-$k$ scaling
behavior of $S(k) \sim k^\alpha$ determines the large-$R$
asymptotic behavior of $\sigma_N^2(R)$, based on which all DHU
systems can be categorized into three classes:
$\sigma_N^2(R) \sim R^{d-1}$ for $\alpha>1$ (class I); $\sigma_N^2(R)
\sim R^{d-1}\ln(R)$ for $\alpha=1$ (class II); and $\sigma_N^2(R)
\sim R^{d-\alpha}$ for $0<\alpha<1$ (class III) \cite{To18a}. Since its original introduction by Torquato and Stillinger \cite{To03}, the concept of hyperuniformity has been generalized to characterize binary heterogeneous materials \cite{Za09}, random scalar field \cite{Ma17, To16a}, interfacial statistics, anisotropic systems and divergence-free vector fields \cite{To16a}.


A wide spectrum of physical and biological systems have been identified to possess the property of hyperuniformity \cite{Ga02, Do05, Za11a, Ji11, Ch14, Za11b, To15, Uc04, Ba08, Ba09, Le83, Zh15a, Zh15b, Ku11, Hu12, Dr15, He15, Ja15, We15, To08, Fe56, Ji14, Ma15, He13, Kl19, Le19, Ch18b, ding2018hyperuniform}. DHU materials are found to possess superior physical properties including large isotropic photonic band gaps \cite{Fl09, Ma13}, optimized transport properties \cite{Zh16, Ch18a}, mechanical properties \cite{Xu17}, wave-propagation characteristics \cite{Ch18a, Kl18, Le16}, as well as optimal multi-functionalities \cite{To18b}. Very recently, DHU patterns of electrons emerging from a quantum jamming transition of correlated many-electron state in 2D materials, which leads to enhanced electronic transport, has been observed \cite{Ge19}. In addition, it is found that DHU distribution of localized electrons in 2D amorphous silica results in an insulator-metal transition in the material \cite{Zh20}. These discoveries suggest the existence of a novel DHU state of electrons in low dimensional materials and shed lights on novel device applications by exploring the emergent properties of the DHU electron states.

A number equilibrium and non-equilibrium pathways to achieving DHU states have also been identified. For example, classical many-body systems with degenerate DHU ground states can be obtained by imposing appropriate inter-particle pair potentials \cite{To15, Zh15a, Zh15b}. Jamming transition in hard particle systems leads to the non-equilibrium maximally random jammed (MRJ) packing states that are class-II hyperuniform \cite{Do05, Za11a, zachary2011hyperuniformity, zachary2011hyperuniformity2, Ji11}. In addition, although individual point defects such as vacancies and intersitials in perfect crystals tend to destroy hyperuniformity \cite{kim2018effect}, it is found that the Stone-Wales defects (corresponding to bonded dislocation pairs) preserve hyperuniformity in 2D materials and can continuously transform the material from the crystalline state to a variety of hyperuniform amorphous state \cite{chen2021stone}.

It is found that certain sets of stochastic displacements in perfect lattices can preserve hyperuniformity, but could lead to hyperuniformity class transitions in the system \cite{Ga04a, Ki18, Kl20}. Klatt et al. \cite{Kl19} demonstrated that one can convert a wide spectrum of nonhyperuniform and hyperuniform point patterns into effectively stealthy hyperuniform patterns via an iterative tessellation and optimization procedure. Kim et al. \cite{Ki19} discovered that one can convert a variety of point patterns into perfectly hyperuniform materials by assigning a sphere of a particle size at each point in the system. This operation shared some similarities with the “equal-volume tessellation” method to generate hyperuniform point patterns \cite{Ga08}. Moreover, DHU states can also be achieved via absorbing transitions \cite{He15}, as well as in active fluids \cite{Le19, lei2019hydrodynamics} and random organizing systems \cite{hexner2017noise, hexner2017enhanced, weijs2017mixing}.








Here we explore another route to achieving DHU states by investigating equilibrium hyperuniform density distributions of Brownian particles induced by external potentials. Specifically, we analytically derive sufficient conditions on the external potential in order to achieve distinct classes of DHU states of Brownian particles in thermal equilibrium, through examining the stationary-state solutions of the corresponding Smoluchowski equation. We demonstrate that a diversity of desirable DHU states of Brownian particles can be controlled and achieved via certain hyperuniform tight-binding potentials. We also find that thermal motions in these systems tend to enhance hyperuniformity. Moreover, we analyze the evolution dynamics of an initial density distribution (hyperuniform or non-hyperuniform) to the desirable equilibrium DHU state driven by prescribed external potentials, and show that the small wave-number behavior of the density distribution is determined by the full spectra of the force fields associated with the imposed potentials. We find that the transient density distribution can rapidly develop local patterns reminiscent of those in the equilibrium distribution, which is governed by the fast dynamics induced by the external potential. However, the overall distribution is still modulated by the initial density fluctuations which are relaxed through slow diffusive dynamics. These results have implications for the fabrication of designer DHU materials for various applications.


The rest of the paper is organized as follows: In Sec. II, we present the definitions of hyperuniform scalar fields and  backgrounds for the dynamics of Brownian particles in a conservative force field. In Sec. III, we introduce the general procedure for devising designer external potential for achieving hyperuniform equilibrium density distributions of Brownian particles and discuss a general class of tight-binding potentials. In Sec. IV, we examine the dynamics that drives a system to different DHU states, focusing on the different roles of diffusion and external forces. In Sec. V, we make concluding remarks.

\section{Hyperuniformity and Smoluchowski Equation}

\subsection{Hyperuniform Random Scalar Fields}






Consider a statistically homogeneous random scalar field $F({\bf x})$ in $d$-dimensional Euclidean space $\mathbb{R}^d$ that is real valued, such as the density distribution of Brownian particles, the associated autocovariance function is defined as \cite{Ma17, To16a}
\begin{equation}
\label{eq_psi} \psi({\bf r}) = <(F({\bf x}_1)-<F({\bf
x}_1)>)(F({\bf x}_2)-<F({\bf x}_2)>)>,
\end{equation}
where ${\bf r} = {\bf x}_2-{\bf x}_1$. The spectral density
function $\hat{\psi}({\bf k})$ is given by
\begin{equation}
\hat{\psi}({\bf k}) = \int_{\mathbb{R}^d} \psi({\bf r})e^{-i{\bf
k}\cdot{\bf r}}d{\bf r},
\end{equation}
which is the Fourier transform of $\psi(r)$. The hyperuniform
condition is then given by
\begin{equation}
\label{eq_hyper} \lim_{|{\bf k}|\rightarrow 0}\hat{\psi}({\bf k})
= 0,
\end{equation}
which implies that
\begin{equation}
\int_{\mathbb{R}^d} \psi({\bf r})d{\bf r} = 0.
\end{equation}

Equivalently, consider a spherical observation window with radius $R$ and volume $v_1(R)$ in $\mathbb{R}^d$. The integrated field within the window
will fluctuate as the window moves in the system. The associated variance $\sigma_F^2(R)$ is given by
\begin{equation}
\sigma_F^2(R) = \frac{1}{v_1(R)}\int_{\mathbb{R}^d} \psi({\bf
r})\alpha(r; R)d{\bf r},
\end{equation}
where $\alpha(r; R)$ is the scaled intersection volume, i.e., the
ratio of the intersection volume of two spherical windows of
radius $R$ whose centers are separated by a distance $r$ to
$v_1(R)$. For a hyperuniform scalar field in $\mathbb{R}^d$,
$\sigma_F^2(R)$ decreases more rapidly than ${R}^d$ for large $R$ \cite{To16a}, i.e.,
\begin{equation}
\lim_{R\rightarrow \infty} R^d \sigma_F^2(R) = 0.
\end{equation}

In certain cases, the overall field $F({\bf x})$
can be considered as a linear superposition of many kernels
$K({\bf x}, {\bf r}_i)$ resulting from source
$i$ centered at ${\bf r}_i$, i.e.,
\begin{equation}
\label{eq_field} F({\bf x}) = \sum_i K({\bf x}, {\bf r}_i).
\end{equation}
It has been shown in Ref. \cite{To16a} that the spectral density
for a scalar field given by Eq. (\ref{eq_field}) can be written as
\begin{equation}
\label{eq_spectral} \hat{\psi}({\bf k}) = \rho \hat{K}^2({\bf
k})S({\bf k}),
\end{equation}
where $\rho$ is the number density associated with the sources, $\hat{K}({\bf
k})$ is the Fourier transform of the kernel field
$K({\bf x})$, and $S({\bf k})$ is the structure factor associated
with the distribution of the source centers.

\subsection{Smoluchowski Equation for Brownian Particles in an External Potential Field}


The dynamics of a Brownian particle in a force field ${\bf f}({\bf r})$ is described by the Langevin equation \cite{langevin1908theorie, uhlenbeck1930theory}, i.e.,
\begin{equation}
m\frac{d^2{\bf r}}{dt^2} = -\gamma \frac{d{\bf r}}{dt} + {\bf f}({\bf r}) + \sigma \boldsymbol{\xi}(t)
\label{eq_Langevin}
\end{equation}
where ${\bf r}$ is the position vector of the particle, $\gamma$ is the scalar friction constant, $\sigma$ is the amplitude of the fluctuating force $\xi$, satisfying $<\boldsymbol{\xi}(t_i) \cdot \boldsymbol{\xi}(t_j) > = \delta(t_i - t_j)$. In the ensuing discussion, we focus on the strong friction limit, i.e., the friction force is much larger than the force of inertia. In this limit, Eq. (\ref{eq_Langevin}) reduces to
\begin{equation}
\gamma \frac{d{\bf r}}{dt} = {\bf f}({\bf r}) + \sigma \boldsymbol{\xi}(t)
\label{eq_L2}
\end{equation}

It is well known that the Fokker-Planck equation for the particle density distribution $P({\bf r}, t)$ associated with Eq. (\ref{eq_L2}) is given by \cite{rosenbluth1957fokker, kadanoff2000statistical}
\begin{equation}
\frac{\partial P({\bf r}, t)}{\partial t} = (\bigtriangledown^2 \frac{\sigma^2}{2\gamma^2}-\bigtriangledown \cdot \frac{\bf {f}({\bf r})}{\gamma})  P({\bf r}, t)
\label{eq_FP}
\end{equation}
Note that we can define the diffusion coefficient $D = \frac{\sigma^2}{2\gamma^2}$, and Eq. (\ref{eq_FP}) becomes
\begin{equation}
\frac{\partial P({\bf r}, t)}{\partial t} = \bigtriangledown \cdot (\bigtriangledown D- \frac{\bf {f}({\bf r})}{\gamma})  P({\bf r}, t)
\label{eq_FP2}
\end{equation}
from which we can define the flux
\begin{equation}
{\bf j}({\bf r}, t) =  (\bigtriangledown D- \frac{\bf {f}({\bf r})}{\gamma})  P({\bf r}, t)
\end{equation}
It can be easily seen that in the case ${\bf f}({\bf r}) = 0$, Eq. (\ref{eq_FP2}) reduces to the normal diffusion equation.

Evoking the fluctuation-dissipation theorem,
\begin{equation}
\bigtriangledown D = {\bf f}({\bf r}) (\gamma^{-1} - D \beta)
\end{equation}
where $\beta = 1/k_B T$ and $k_B$ is the Boltzmann constant, which reduces to $\sigma^2 = 2k_B T \gamma$ in the case of spatially invariant $D$, we can obtain the Smoluchowski equation \cite{smoluchowski1927drei, melzak1957scalar}, i.e.,
\begin{equation}
\frac{\partial P({\bf r}, t)}{\partial t} = \bigtriangledown \cdot D (\bigtriangledown - \beta {\bf f}({\bf r}))  P({\bf r}, t)
\label{eq_Sm}
\end{equation}
If we further consider the force field is resulted from a scalar potential field $U({\bf r})$, i.e., ${\bf f}({\bf r}) = - \bigtriangledown U({\bf r})$, the Smoluchowski equation (\ref{eq_Sm}) can be written as
\begin{equation}
\frac{\partial P({\bf r}, t)}{\partial t} = \bigtriangledown \cdot D e^{-\beta U({\bf r})} \bigtriangledown e^{\beta U({\bf r})} P({\bf r}, t)
\label{eq_Sm2}
\end{equation}
It follows immediately from Eq. (\ref{eq_Sm2}) that
\begin{equation}
P({\bf r}) \sim e^{-\beta U({\bf r})}
\label{eq_sol}
\end{equation}
provides a steady-state solution, which also leads to
\begin{equation}
{\bf j}({\bf r}, t) = D e^{-\beta U({\bf r})} \bigtriangledown e^{\beta U({\bf r})} P({\bf r})  = 0.
\end{equation}
This indicates that Eq. (\ref{eq_sol}) is also the {\it equilibrium} density distribution of Brownian particles in the external potential field $U({\bf r})$.




\section{Designer External Potential for Hyperuniform Equilibrium Density Distributions of Brownian Particles}

\subsection{Designer Potential from Equilibrium Solution of Smoluchowski Equation}




In the ensuing discussions, we will mainly focus on 1D systems, which allow insightful and illustrative analytical treatments. However, we note the general procedures introduced here and the insights obtained are readily applicable in other dimensions. In particular, consider the 1D system of $N$ point-like Brownian particles in a domain of length $L$ with a number density $\rho_0 = N/L$, which is also well defined in the thermodynamic limit $N, L \rightarrow \infty$. In the presence of an external potential field $U(r)$, the equilibrium density distribution is given by Eq. (\ref{eq_sol}), i.e.,
\begin{equation}
P(r) = \rho(\beta) e^{-\beta U(r)}
\label{eq_P}
\end{equation}
where $\rho(\beta)$ is a temperature dependent normalization factor determined by
\begin{equation}
\rho(\beta) \int_0^L e^{-\beta U(r)}dr = N
\end{equation}
In the case $U(r) = 0$, it is clear that $\rho(\beta) = \rho_0$.

It follows immediately from Eq. (\ref{eq_P}) that
\begin{equation}
U(r) = - \beta^{-1} \ln[P(r)/\rho(\beta)]
\label{eq_Ur}
\end{equation}
which is the required external potential field to achieve a target equilibrium particle density distribution at $T = (k_B \beta)^{-1}$. Therefore, if $P(r)$ is a hyperuniform density distribution, Eq. (\ref{eq_Ur}) allows one to obtain the corresponding $U(r)$ to achieve the hyperuniform distribution.



We note that Eq. (\ref{eq_Ur}) indicates that the temperature $T$ (or equivalently $\beta$) plays a nontrivial role in determining the potential field for a given density distribution. In particular, to achieve the same distribution at different temperatures, different potential fields are required to suppress thermal fluctuations at different levels. On the other hand, for a given potential field, the corresponding equilibrium density distribution at different $T$ is also different. For example, at very high temperature ($T \gg 1$), $P(r) \sim \rho_0$ is an almost flat distribution regardless of the functional form of $U(r)$. These temperature effects will be further illustrated and discussed in detail using numerical examples in the ensuing sections.

\subsection{Applications to Tight-Binding Potentials}

We now focus on a special class of potentials, i.e., the tight-binding potentials which can be written as
\begin{equation}
U(r) = \sum_{i = 1}^M \phi(r-r_i)
\label{eq_Ur2}
\end{equation}
where $\phi(r)$ is the (localized) potential resulted from a source located at the origin, and $r_i$ is the location of the $i$th source. In classical systems, the tight-binding potentials can be realized by placing ``attractors'' or ``repellers'' at designated locations. In quantum systems, Eq. (\ref{eq_Ur2}) is typically used in linear combination of orbital approximations to treat localized electron states in solid materials.

Substituting Eq. (\ref{eq_Ur2}) into Eq. (\ref{eq_P}) yields
\begin{equation}
P(r) =  \rho(\beta) \prod_{i = 1}^M e^{-\beta \phi(r-r_i)} = \rho(\beta) \prod_{i = 1}^M P_i(r)
\end{equation}
where
\begin{equation}
P_i(r) = e^{-\beta \phi(r-r_i)}
\label{eq_Pi}
\end{equation}
which indicates that the equilibrium density distribution resulted from a tight-binding potential is the product of local density distributions $P_i(r)$ resulted from individual potential $\phi(r-r_i)$. In the following discussion, we consider two commonly used forms of $\phi(r)$, i.e., the square potential and the Gaussian potential.

\subsubsection{Square Potential}

\begin{figure}
\includegraphics[width=0.375\textwidth,keepaspectratio]{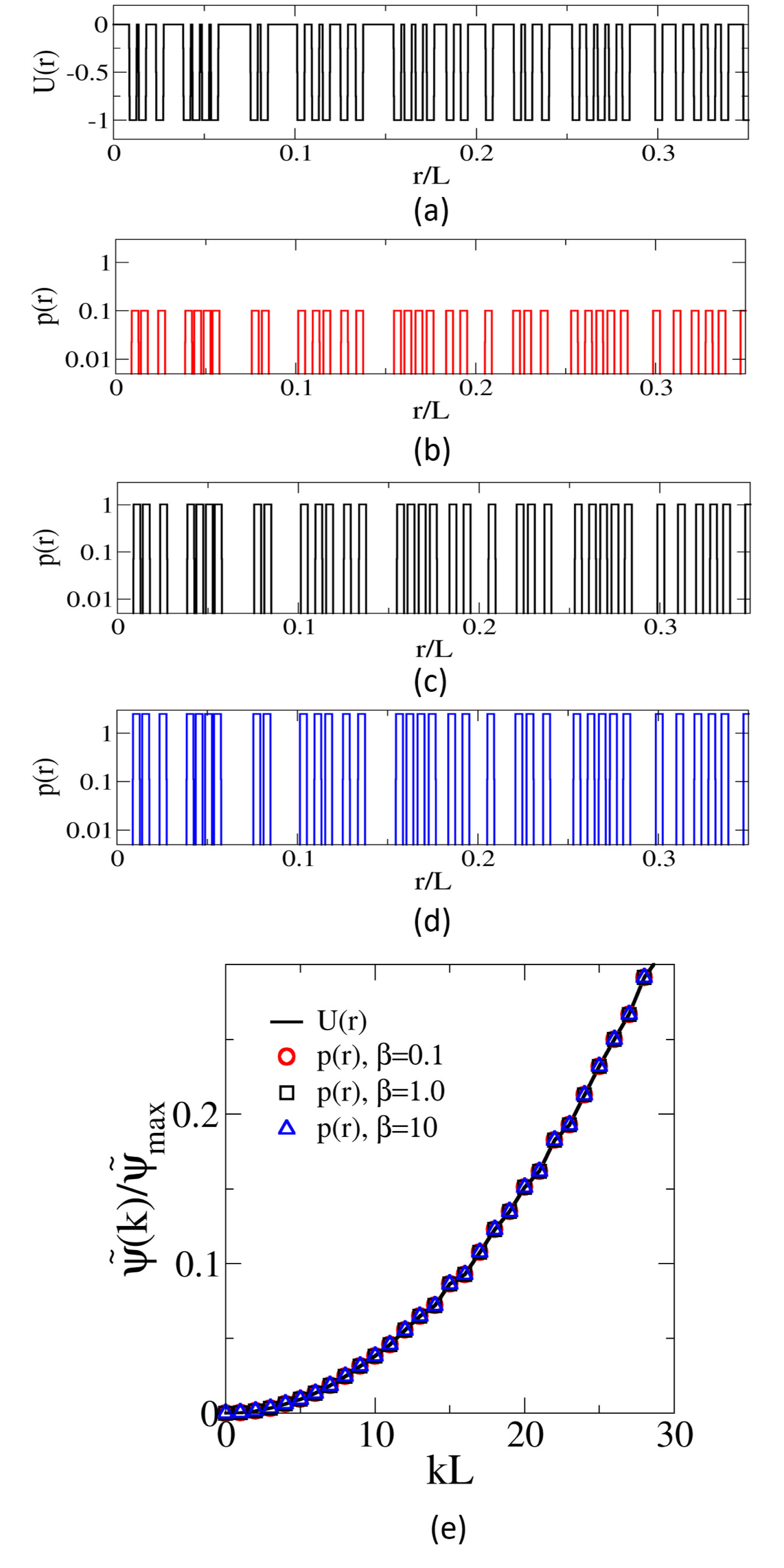}
\caption{Designer tight-binding potential based on the square-well potentials (a) and the resulting hyperuniform equilibrium density distributions of Brownian particles at different temperatures (i.e., $\beta = 0.1$, 1 and 10, (b)-(d)). The distribution of the potential centers is class-I hyperuniform, and possess a structure factor $S(k) \sim k^2$ for small $k$. In the numerical examples, we have used $M = 100$, $U_0 = 1$, $\sigma = 0.002 L$ (the distance between two closest potential centers is $0.0025L$), and $\rho_0 = 1$. In this case, the density distribution should be interpreted as the probability of finding a particle in a specific location within the domain. For better visualization, we only show a portion $[0, 0.35L]$ of the entire domain. (e) Spectral densities $\hat{\psi}(k)$ (normalized with respect to the largest value $\hat{\psi}_{max}$) associated with particle density distributions $p(r)$ derived from designer tight-binding potential $U(r)$.} \label{fig_1}
\end{figure}

\begin{figure}
\centering
\includegraphics[width=0.375\textwidth,keepaspectratio]{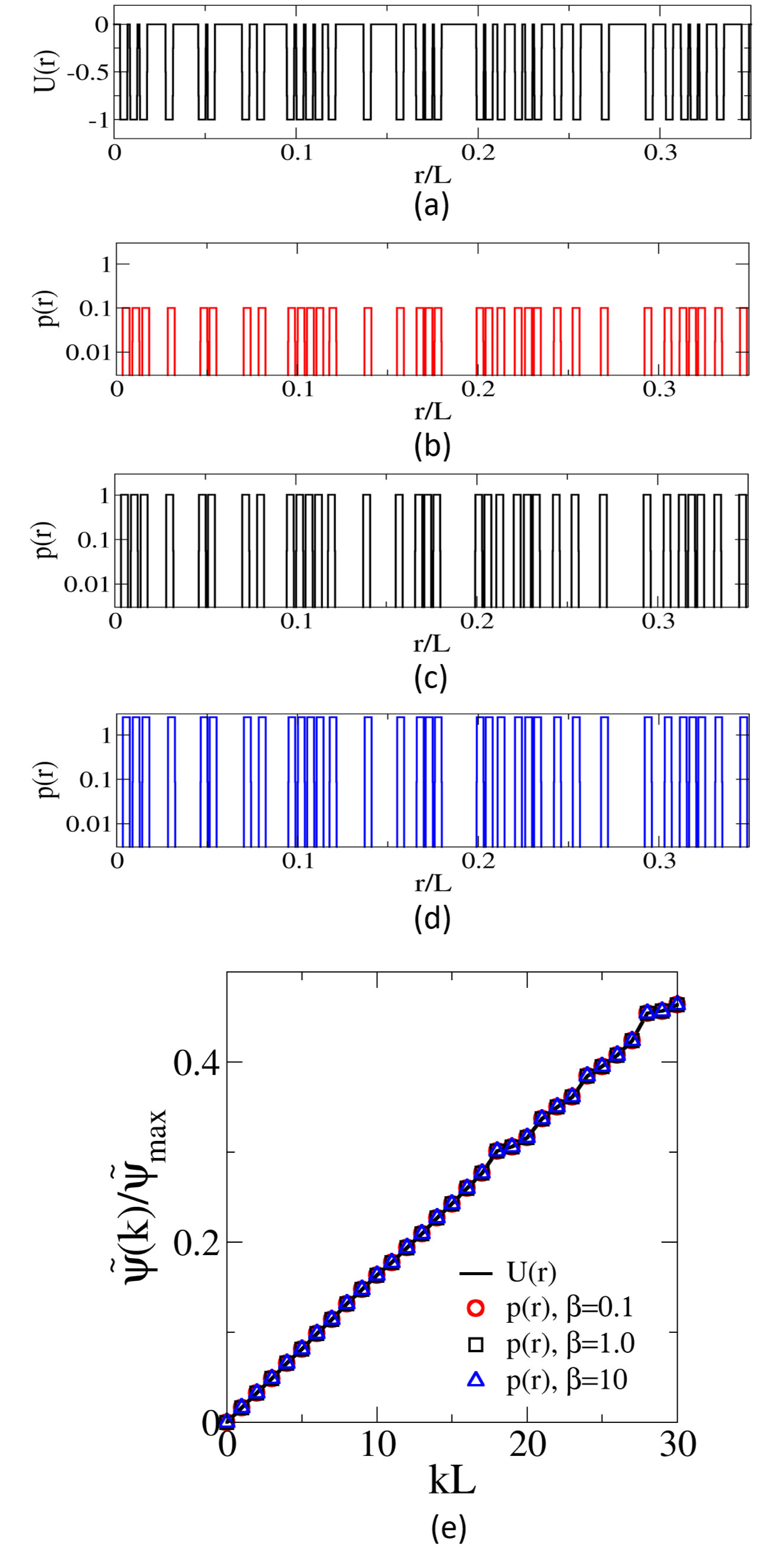}
\caption{Designer tight-binding potential based on the square-well potentials (a) and the resulting hyperuniform equilibrium density distributions of Brownian particles at different temperatures (i.e., $\beta = 0.1$, 1 and 10, (b)-(d)). The distribution of the potential centers is class-II hyperuniform, and possess a structure factor $S(k) \sim k$ for small $k$. In the numerical examples, we have used $M = 100$, $U_0 = 1$, $\sigma = 0.002 L$ (the distance between two closest potential centers is $0.0025L$), and $\rho_0 = 1$. In this case, the density distribution should be interpreted as the probability of finding a particle in a specific location within the domain. For better visualization, we only show a portion $[0, 0.35L]$ of the entire domain. (e) Spectral densities $\hat{\psi}(k)$ (normalized with respect to the largest value $\hat{\psi}_{max}$) associated with particle density distributions $p(r)$ derived from designer tight-binding potential $U(r)$.} \label{fig_2}
\end{figure}

The square potential is defined as
\begin{equation}
\phi(r) = \left\{{\begin{array}{cc} -U_0, &
|r| \le \sigma
 \\\\ 0, & |r|>\sigma.
\end{array}}\right.
\label{eq_SQP}
\end{equation}
where $U_0$ and $2\sigma$ are respectively the depth and the width of the square well. Substituting Eq. (\ref{eq_SQP}) into Eq. (\ref{eq_Pi}) yields
\begin{equation}
P_i(r) = \left\{{\begin{array}{cc} e^{\beta U_0}, &
|r-r_i| \le \sigma
 \\\\ 1, & |r-r_i|>\sigma.
\end{array}}\right.
\end{equation}
It can be easily seen that if the separation distance between the sources are larger than the square width, i.e.,
\begin{equation}
\min_{i<j}\{|r_i - r_j|\} \ge 2\sigma
\label{eq_dist}
\end{equation}
the overall particle density distribution is then given by
\begin{equation}
P(r) = \rho(\beta)\left \{\sum_{i=1}^M (e^{\beta U_0}-1) \Theta(r-r_i) + 1\right \}
\label{eq_PS}
\end{equation}
where
\begin{equation}
\Theta(r) = \left\{{\begin{array}{cc} 1, &
|r-r_i| \le \sigma
 \\\\ 0, & |r-r_i|>\sigma.
\end{array}}\right.
\label{eq_theta}
\end{equation}
is the indicator function, and
\begin{equation}
\rho(\beta) = N \left [ L + (e^{\beta U_0} - 1) 2M\sigma  \right ]^{-1}
\end{equation}

Without loss of generality, we focus on the nontrivial fluctuating part of $P(r)$ due to the potential field, i.e.,
\begin{equation}
p(r) = P(r)-\rho(\beta) = \rho(\beta) (e^{\beta U_0}-1) \sum_{i=1}^M  \Theta(r-r_i)
\end{equation}
It follows immediately from Eq. (\ref{eq_field}) and Eq. (\ref{eq_spectral}) that
\begin{equation}
K(r) = \rho(\beta) (e^{\beta U_0}-1) \Theta(r)
\end{equation}
whose Fourier transform is given by
\begin{equation}
\hat{K}(k) = 2\rho(\beta) (e^{\beta U_0}-1) \frac{\sin(k\sigma)}{k}
\label{eq_Kk}
\end{equation}
Therefore, the spectral density $\psi$ associated with $p(r)$ is given by
\begin{equation}
\hat{\psi}(k) = 4 \rho_s \left [ \rho(\beta) (e^{\beta U_0}-1) \frac{\sin(k\sigma)}{k} \right ]^2 S(k)
\end{equation}
where $\rho_s = M/L$ is the number density associated with square potentials, and $S(k)$ is the structure factor associated with the distribution of square potential centers $\{r_i\}$. At $k \rightarrow 0$ limit, we have
\begin{equation}
\hat{\psi}(k) \approx 4 \rho_s \left [ \rho(\beta) (e^{\beta U_0}-1) \sigma \right ]^2 S(k)
\label{eq_psi0}
\end{equation}
where we have ignore higher-order terms in the Taylor expansion of $\sin(k\sigma)/k$ around $k = 0$.

It can be clearly seen from Eq. (\ref{eq_psi0}) that at a given temperature (i.e., $\beta$), the small-$k$ behavior of $\hat{\psi}(k)$ is determined by that of $S(k)$. This implies that a potential field $U(r)$ composed of hyperuniformly arranged square potentials satisfying Eq. (\ref{eq_dist}) can always lead to a hyperuniform density distribution $p(r)$, with the same hyperuniformity class.


It is interesting to note the system can achieve hyperuniformity regardless of the properties of $U(r)$ (e.g., the distribution $\{r_i\}$ and small-$k$ behavior of $S(k)$) if the following equation holds:
\begin{equation}
e^{\beta U_0} - 1 = 0
\end{equation}
We note this can be achieved under two conditions: (i) $U_0 = 0$, which leads to the trivial situation of absence of external potentials; and (ii) $\beta = 0$ (i.e., $T \rightarrow \infty$). This implies that thermal fluctuations, which grow stronger as $T$ increases, enhance hyperuniformity in the system. At this limit, $P(r) = \rho$ is a location independent constant with $p(r) = 0$.



\begin{figure}[H]
\centering
\includegraphics[width=0.375\textwidth,keepaspectratio]{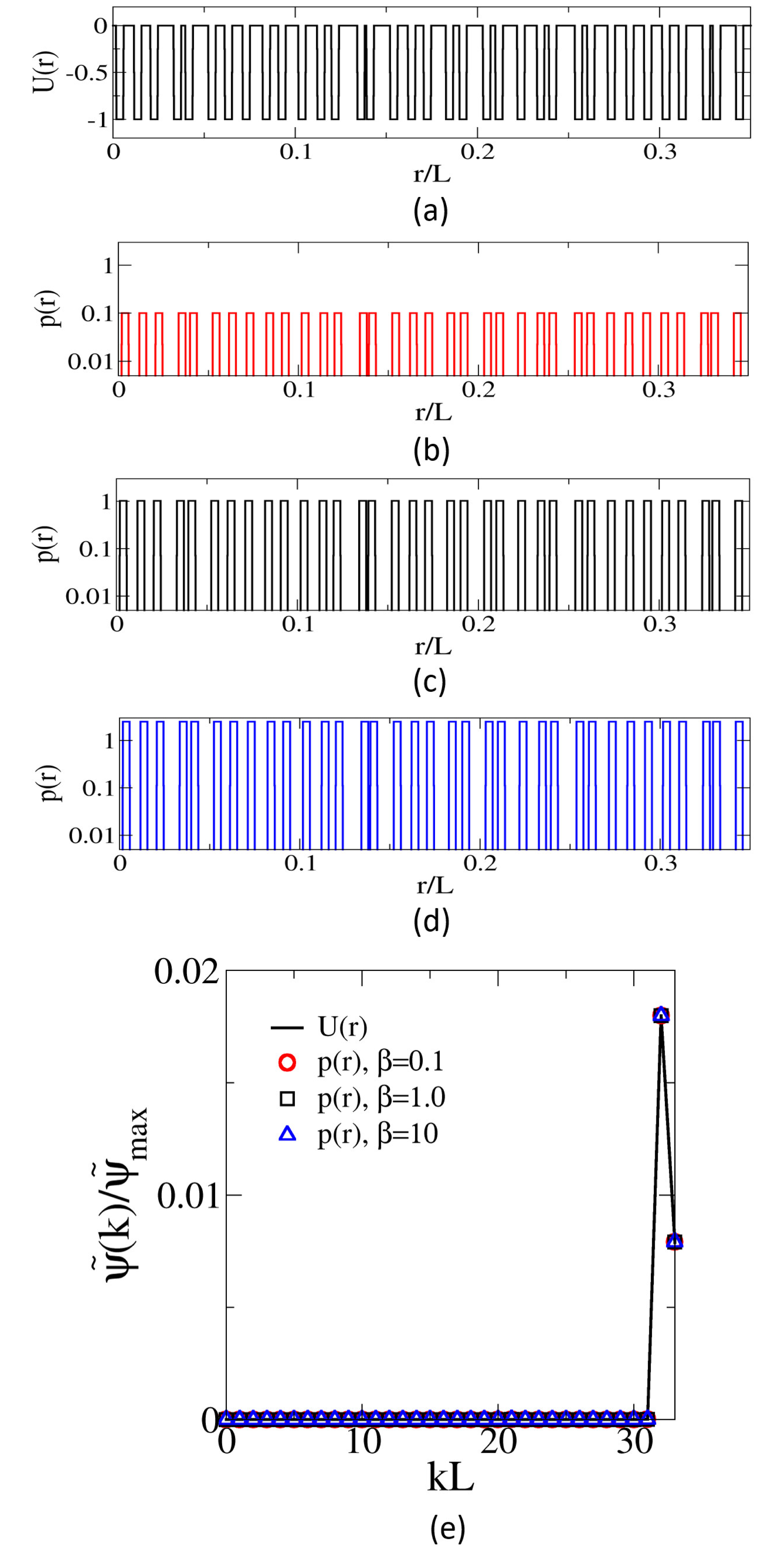}
\caption{Designer tight-binding potential based on the square-well potentials (a) and the resulting hyperuniform equilibrium density distributions of Brownian particles at different temperatures (i.e., $\beta = 0.1$, 1 and 10, (b)-(d)). The distribution of the potential centers is stealthy hyperuniform, and possess a structure factor $S(k) = 0$ for $k<K^*$. In the numerical examples, we have used $M = 100$, $U_0 = 1$, $\sigma = 0.002 L$ (the distance between two closest potential centers is $0.0025L$), and $\rho_0 = 1$. In this case, the density distribution should be interpreted as the probability of finding a particle in a specific location within the domain. For better visualization, we only show a portion $[0, 0.35L]$ of the entire domain. (e) Spectral densities $\hat{\psi}(k)$ (normalized with respect to the largest value $\hat{\psi}_{max}$) associated with particle density distributions $p(r)$ derived from designer tight-binding potential $U(r)$.} \label{fig_3}
\end{figure}

Figures \ref{fig_1}, \ref{fig_2} and \ref{fig_3} show examples of $U(r)$ composed of hyperuniform distributions of square potentials with different hyperuniformity classes, and the resulting hyperuniform density distributions at different $\beta$, as well as the associated spectral densities normalized with respect to the corresponding maximal values (i.e., $\hat{\psi}(k)/\hat{\psi}_{max}$). The potential fields $U(r)$ are constructed by numerically generating hyperuniform distributions of the square potential located at $\{r_i\}$. Specifically, a generalized ``collective coordinate'' method was employed \cite{ding2018hyperuniform, Ba08}, which stochastically optimizes the distribution while satisfying the exclusion constraints imposed by Eq. (\ref{eq_dist}). We generated representative distributions associated with $S(k) \sim k^2$ (class-I hyperuniformity), $S(k) \sim k$ (class-II hyperuniformity), and $S(k) = 0$ for $k \le K^*$ (stealthy distributions). It can be clearly seen from the computed spectral densities (shown in Fig. 1-3e, respectively) that the hyperuniformity of $U(r)$ is preserved in the density distributions. In particular, the normalized spectral densities are virtually indistinguishable from one another. Moreover, as $\beta$ decreases ($T$ increases), the fluctuations in all of the resulting density distributions rapidly diminish, indicating the distributions are approaching the perfectly uniform limit.



\subsubsection{Gaussian Potential}

\begin{figure}[H]
\includegraphics[width=0.375\textwidth,keepaspectratio]{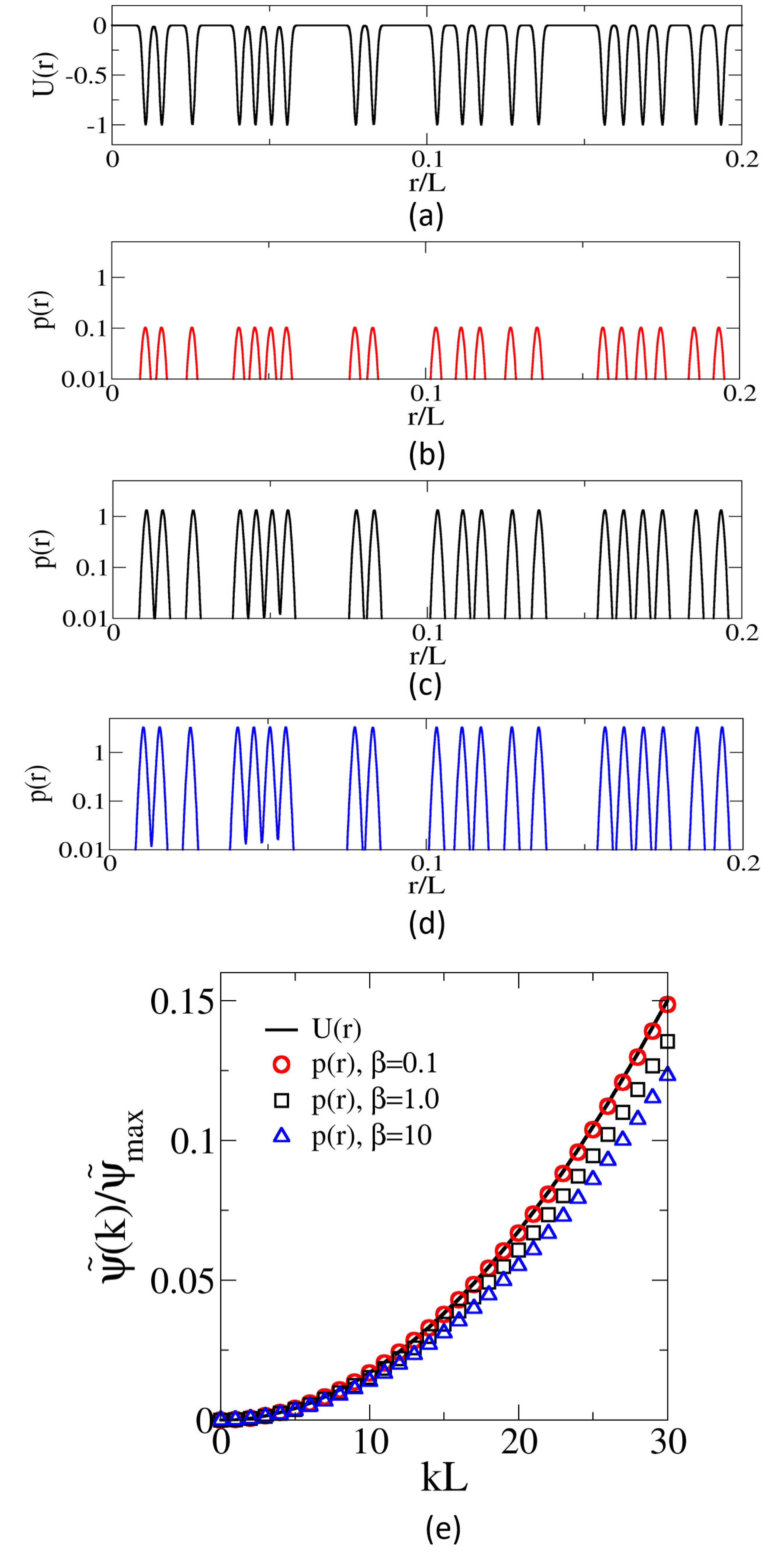}
\caption{Designer tight-binding potential based on the Gaussian potentials (a) and the resulting hyperuniform equilibrium density distributions of Brownian particles at different temperatures (i.e., $\beta = 0.1$, 1 and 2, (b)-(d)). The distribution of the potential centers is class-I hyperuniform, and possess a structure factor $S(k) \sim k^2$ for small $k$. In the numerical examples, we have used $M = 100$, $U_0 = 1$, $\sigma = 0.0008 L$ (the distance between two closest potential centers is $0.0025L$), and $\rho_0 = 1$. In this case, the density distribution should be interpreted as the probability of finding a particle in a specific location within the domain. For better visualization, we only show a portion $[0, 0.2L]$ of the entire domain. (e) Spectral densities $\hat{\psi}(k)$ (normalized with respect to the largest value $\hat{\psi}_{max}$) associated with particle density distributions $p(r)$ derived from designer tight-binding potential $U(r)$.} \label{fig_4}
\end{figure}

\begin{figure}[ht]
\includegraphics[width=0.375\textwidth,keepaspectratio]{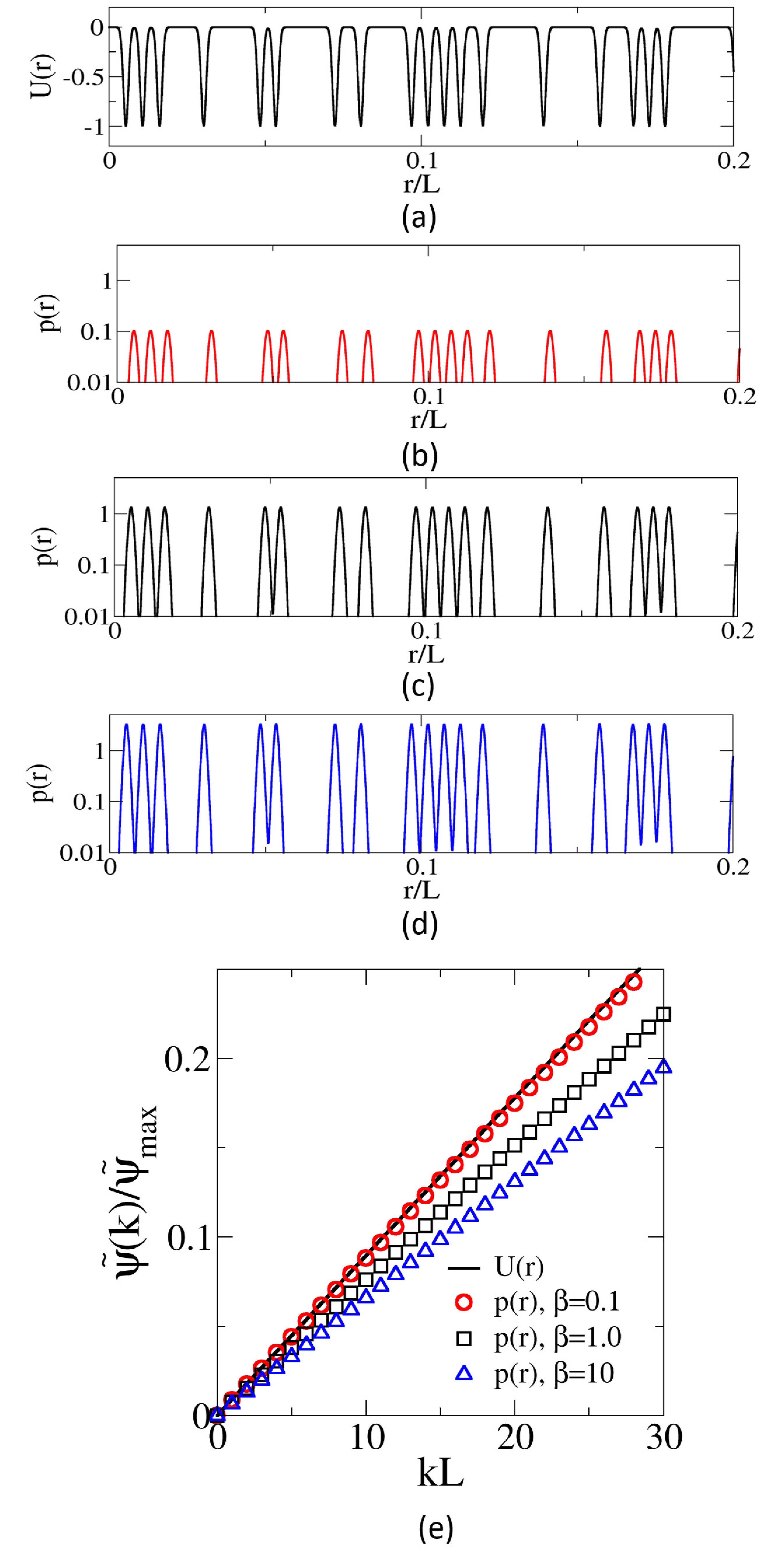}
\caption{Designer tight-binding potential based on the Gaussian potentials (a) and the resulting hyperuniform equilibrium density distributions of Brownian particles at different temperatures (i.e., $\beta = 0.1$, 1 and 2, (b)-(d)). The distribution of the potential centers is class-II hyperuniform, and possess a structure factor $S(k) \sim k$ for small $k$. In the numerical examples, we have used $M = 100$, $U_0 = 1$, $\sigma = 0.0008 L$ (the distance between two closest potential centers is $0.0025L$), and $\rho_0 = 1$. In this case, the density distribution should be interpreted as the probability of finding a particle in a specific location within the domain. For better visualization, we only show a portion $[0, 0.2L]$ of the entire domain. (e) Spectral densities $\hat{\psi}(k)$ (normalized with respect to the largest value $\hat{\psi}_{max}$) associated with particle density distributions $p(r)$ derived from designer tight-binding potential $U(r)$.} \label{fig_5}
\end{figure}

\begin{figure}[ht]
\includegraphics[width=0.375\textwidth,keepaspectratio]{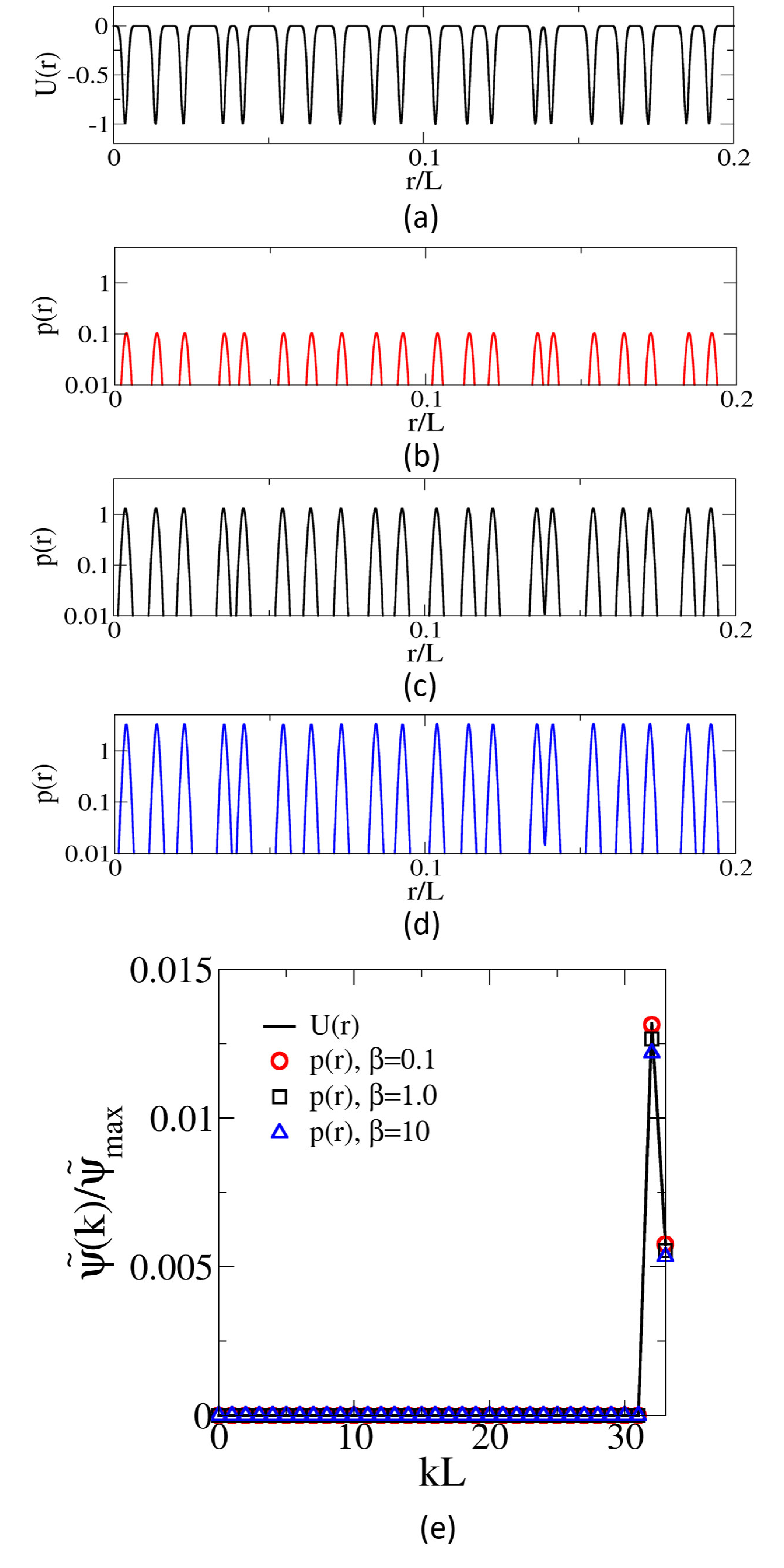}
\caption{Designer tight-binding potential based on the Gaussian potentials (a) and the resulting hyperuniform equilibrium density distributions of Brownian particles at different temperatures (i.e., $\beta = 0.1$, 1 and 2, (b)-(d)). The distribution of the potential centers is stealthy hyperuniform, and possess a structure factor $S(k) = 0$ for $k<K^*$. In the numerical examples, we have used $M = 100$, $U_0 = 1$, $\sigma = 0.0008 L$ (the distance between two closest potential centers is $0.0025L$), and $\rho_0 = 1$. In this case, the density distribution should be interpreted as the probability of finding a particle in a specific location within the domain. For better visualization, we only show a portion $[0, 0.2L]$ of the entire domain. (e) Spectral densities $\hat{\psi}(k)$ (normalized with respect to the largest value $\hat{\psi}_{max}$) associated with particle density distributions $p(r)$ derived from designer tight-binding potential $U(r)$.} \label{fig_6}
\end{figure}

We now consider the Gaussian potential, which is defined as
\begin{equation}
\phi(r) = - U_0 e^{-\frac{r^2}{2\sigma^2}}
\label{eq_G}
\end{equation}
where $U_0$ and $\sigma$ respectively reflects the effective ``depth'' and ``width'' of the potential. Substituting Eq. (\ref{eq_G}) into Eq. (\ref{eq_Pi}) yields
\begin{equation}
P_i(r) = \exp\left (\beta U_0 e^{-\frac{(r-r_i)^2}{2\sigma^2}} \right )
\end{equation}
and the associated particle density distribution is given by
\begin{equation}
P(r) = \rho(\beta) \prod_{i=1}^M \exp\left (\beta U_0 e^{-\frac{(r-r_i)^2}{2\sigma^2}} \right )
\label{eq_PrG}
\end{equation}
We note that Eq. (\ref{eq_PrG}) cannot be rigorously written as a summation in general. However, if the individual Gaussian potential is sufficiently localized compared to their separation distances, e.g.,
\begin{equation}
\min_{i<j}\{|r_i - r_j|\} \ge n\sigma
\label{eq_dist}
\end{equation}
where $n$ is an empirical constant (e.g., $n = 6$), Eq. (\ref{eq_PrG}) can be approximated as
\begin{equation}
P(r) \approx \rho(\beta)\left \{\sum_{i=1}^M \left [\exp\left (\beta U_0 e^{-\frac{(r-r_i)^2}{2\sigma^2}} \right )   -1 \right ] \Theta(r-r_i) + 1\right \}
\label{eq_PG}
\end{equation}
where $\Theta(r)$ is the indicator function defined by Eq. (\ref{eq_theta}). Following the same procedure used for the square potentials, we focus on the nontrivial fluctuating part in the density distribution due to the potential field, i.e.,
\begin{equation}
p(r) \approx \rho(\beta) \sum_{i=1}^M \left [\exp\left (\beta U_0 e^{-\frac{(r-r_i)^2}{2\sigma^2}} \right )   -1 \right ] \Theta(r-r_i)
\label{eq_pG}
\end{equation}
It is clearly that Eq. (\ref{eq_pG}) is the form of Eq. (\ref{eq_field}), with
\begin{equation}
K(r) = \rho(\beta) \left [\exp\left (\beta U_0 e^{-\frac{r^2}{2\sigma^2}} \right )   -1 \right ] \Theta(r)
\end{equation}
We note that in general a closed analytical form of the Fourier transform of the above $K(r)$ is difficult to obtain. Nonetheless, the zero-$k$ limit $\hat K(k \rightarrow 0)$ is given by
\begin{equation}
{\hat K}(0) = \rho(\beta) \int_{-\frac{n\sigma}{2}}^{\frac{n\sigma}{2}} \left [\exp\left (\beta U_0 e^{-\frac{r^2}{2\sigma^2}} \right )   -1 \right ]dr
\label{eq_integral}
\end{equation}
It is clear that for $r \in [-\frac{n\sigma}{2}, \frac{n\sigma}{2}]$,
\begin{equation}
\exp\left (\beta U_0 e^{-\frac{r^2}{2\sigma^2}} \right ) \le \exp\left (\beta U_0 \right )
\end{equation}
Therefore, we have
\begin{equation}
{\hat K}(0) \le \rho(\beta) (e^{\beta U_0}-1) n\sigma
\end{equation}
which is a bounded constant. This indicates that the small wave-number behavior of the associated spectral density of $p(r)$
\begin{equation}
{\hat \psi}(k) \approx \rho_s [{\hat K}^2(0) + O(k^2)] S(k)
\label{eq_psiG}
\end{equation}
is determined by the behavior of $S(k)$ associated with the distribution of the Gaussian potentials, similar to the case of the square potential field. Moreover, it follows immediately from Eq. (\ref{eq_integral}) that in the cases $\beta = 0 $ (i.e., infinite temperature) or $U_0 = 0$ (i.e., absence of external potential), we have ${\hat K}(0) = 0$, which indicates $p(r) = 0$ and $P(r) = \rho$.



Figures \ref{fig_4}, \ref{fig_5} and \ref{fig_6} show examples of $U(r)$ composed of hyperuniform distributions of Gaussian potentials with different hyperuniformity classes as for the square potential case, and the resulting hyperuniform density distributions at different $\beta$. It can be seen that the hyperuniformity of $U(r)$ is again preserved in the density distributions (verified via spectral density calculations shown also in the figures). As $\beta$ decreases ($T$ increases), the fluctuations in the resulting density distributions rapidly diminish as in the square potential cases, indicating the distributions are approaching the perfectly uniform limit.


These two examples (i.e., the square potential field and Gaussian potential field) illustrate that sufficiently localized tight-binding potentials possessing a hyperuniform distribution can lead to hyperuniform density distributions with the same hyperuniformity class. Such hyperuniform potential fields can be experimentally realized using, e.g., optical traps for colloidal systems. In principle, one can achieve an arbitrary hyperuniform density distribution $P(r)$ via a designer potential field $U(r)$ given by Eq. (\ref{eq_Ur}). However, experimentally realizing the desirable potential fields might be challenging.








\section{Dynamics to Hyperuniform States}

\begin{figure}[ht]
\includegraphics[width=0.485\textwidth,keepaspectratio]{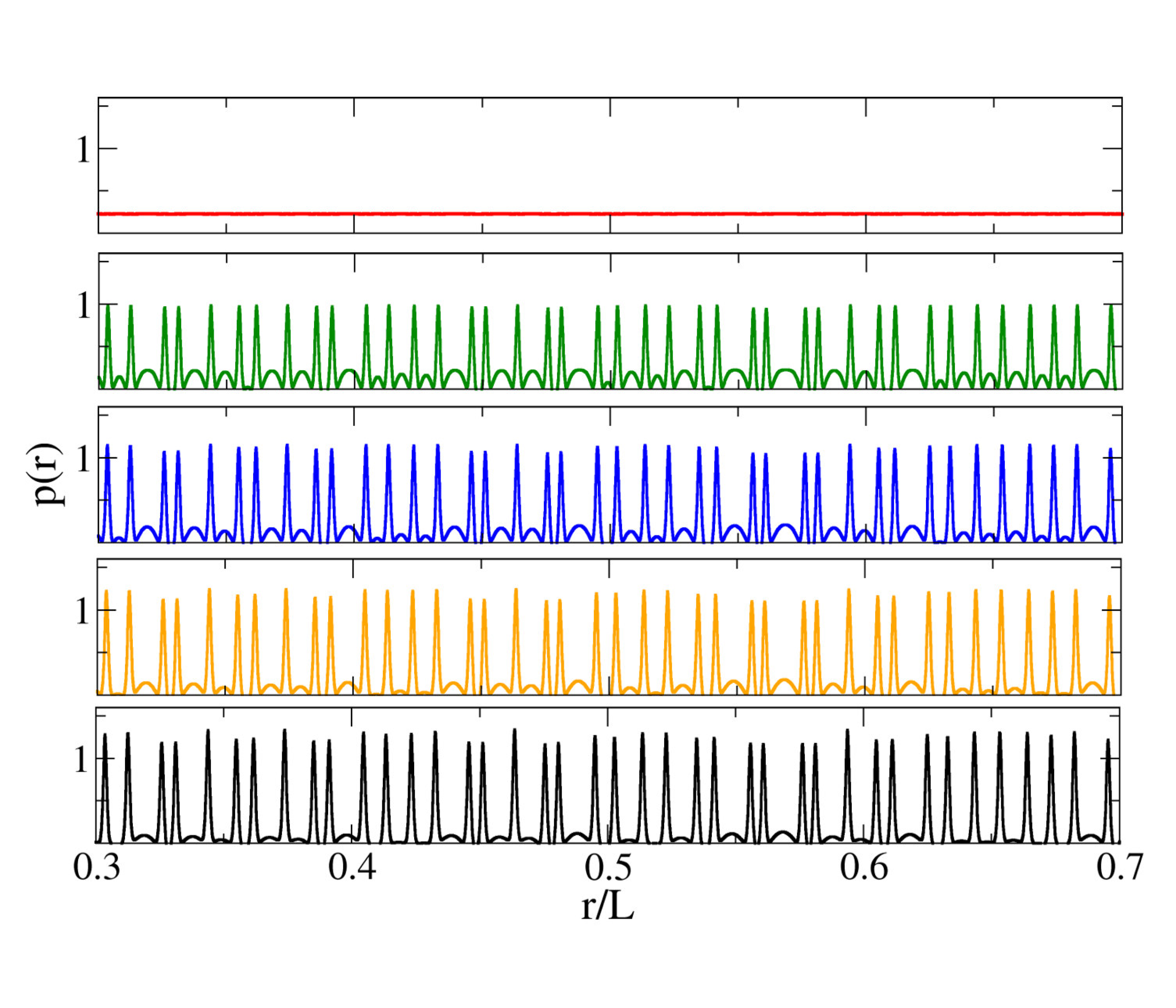}
\caption{Evolution of density distribution $p(r, t) = P(r, t) -
\rho(\beta)$ from a uniform initial distribution $P(r, t = 0) =
1.0$, which is mainly driven by the forces resulted from the
tight-binding Gaussian potential field that is stealthy
hyperuniform. The distribution rapidly converges to the
corresponding equilibrium DHU distribution investigated in Sec.
III. The time steps for the snapshots from top to bottom are
respectively $t = 1$, 1000, 2000, 3000, 5000, where each time step
is $dt = 10^{-8}L^2/D$.} \label{fig_7}
\end{figure}

\begin{figure}[ht]
\includegraphics[width=0.385\textwidth,keepaspectratio]{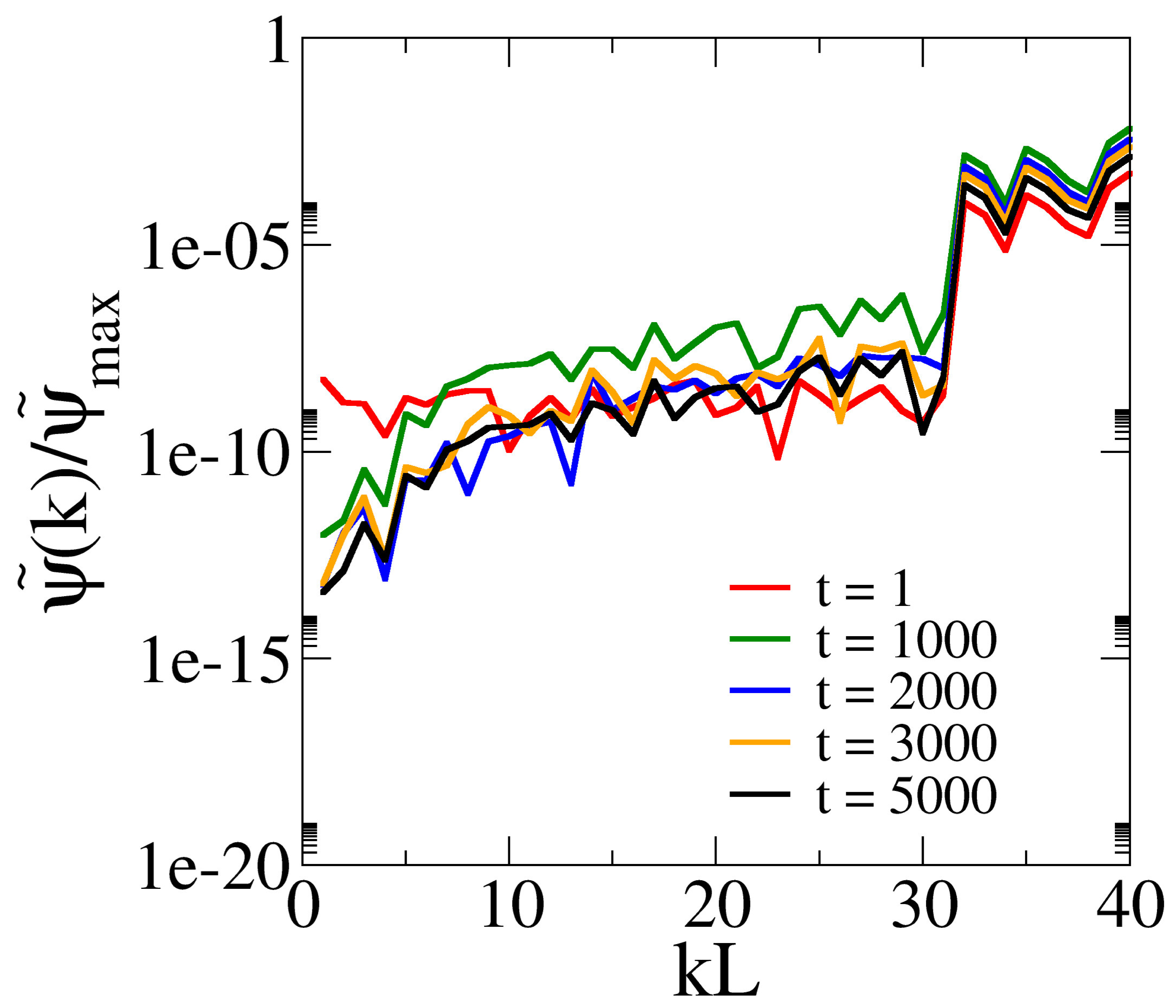}
\caption{Evolution of spectral density $\hat{\psi}(k, t)$
associated with the density distribution $p(r, t) = P(r, t) -
\rho(\beta)$ shown in Fig. \ref{fig_7}.} \label{fig_8}
\end{figure}

\begin{figure}[ht]
\includegraphics[width=0.485\textwidth,keepaspectratio]{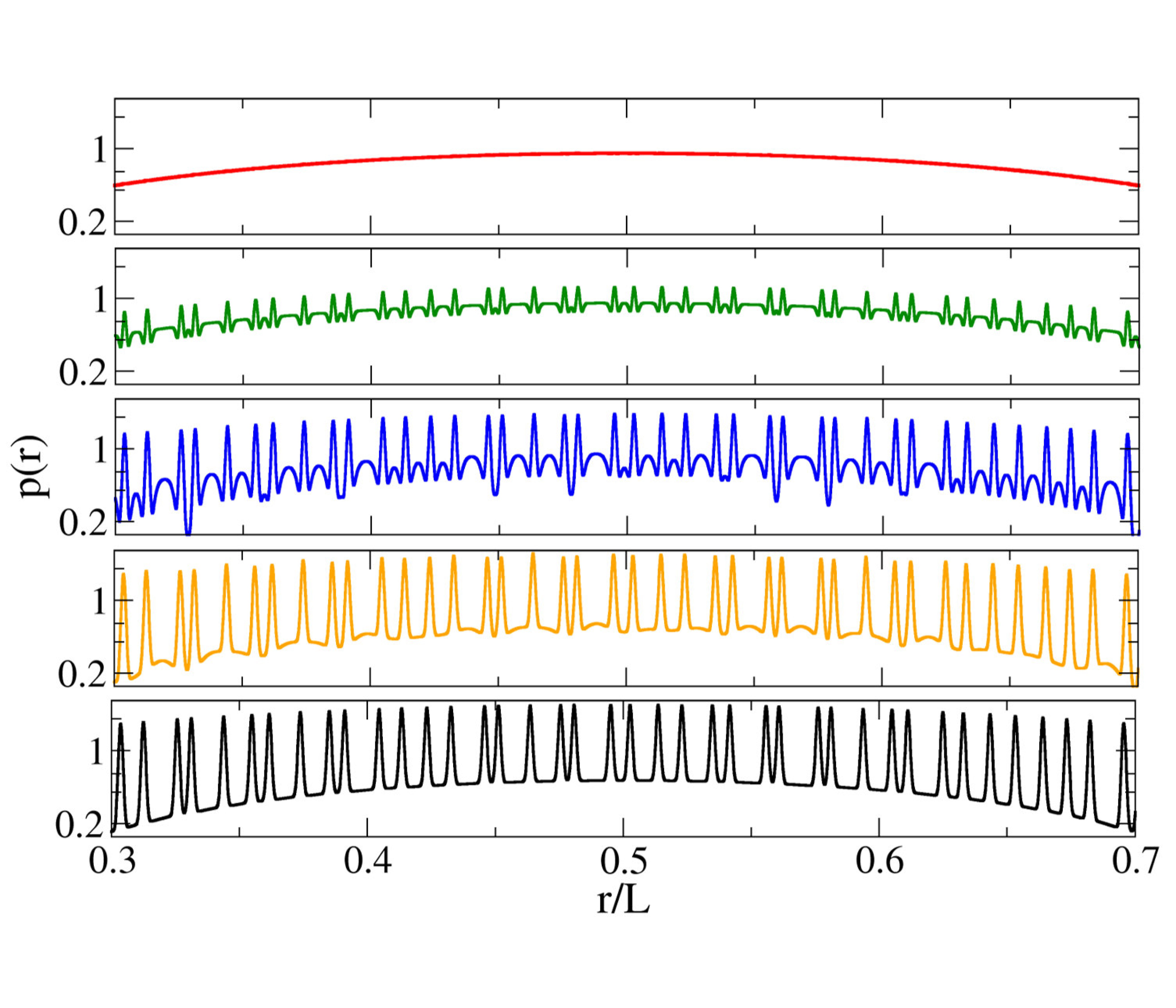}
\caption{Evolution of density distribution $p(r, t) = P(r, t) -
\rho(\beta)$ from a uniform initial distribution $P(r, t = 0) =
\rho_0 \exp[-(r-L/2)^2/a^2]$ (with $a = 0.25L$). Local patterns
reminiscent of the those in the equilibrium DHU distribution are
quickly developed, which are driven by the forces resulted from
the tight-binding Gaussian potential field that is stealthy
hyperuniform. The overall distribution is still modulated by the
initial Gaussian distribution, which slowly relaxes via diffusive
dynamics. The distribution eventually converges to the
corresponding equilibrium DHU distribution investigated in Sec.
III. The time steps for the snapshots from top to bottom are
respectively $t = 1$, 100, 1000, 10000, 100000, where each time
step is $dt = 10^{-8}L^2/D$.} \label{fig_9}
\end{figure}

\begin{figure}[ht]
\includegraphics[width=0.385\textwidth,keepaspectratio]{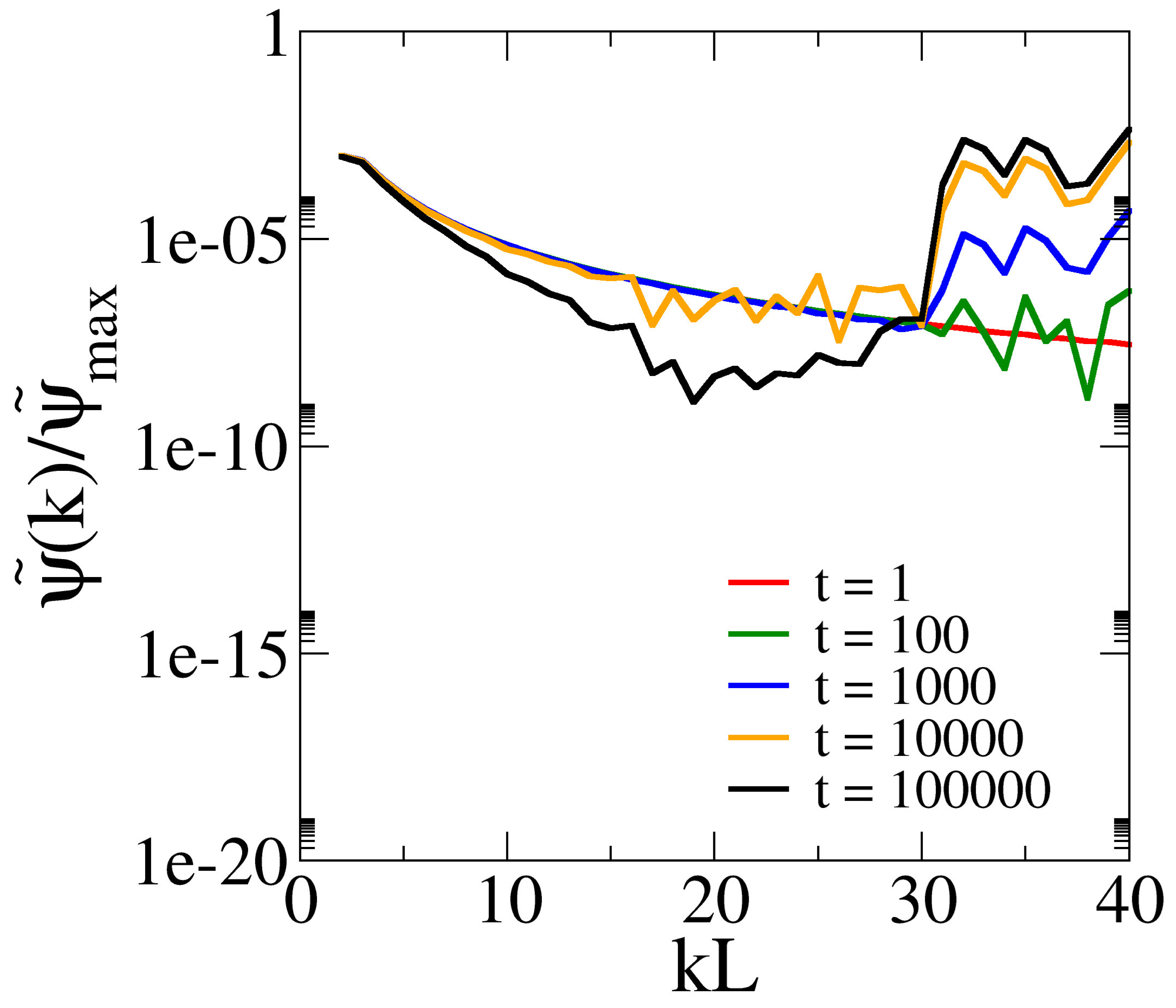}
\caption{Evolution of spectral density $\hat{\psi}(k, t)$
associated with the density distribution $p(r, t) = P(r, t) -
\rho(\beta)$ shown in Fig. \ref{fig_9}.} \label{fig_10}
\end{figure}


We now focus on the evolution dynamics of an initial density distribution (hyperuniform or non-hyperuniform) to the final equilibrium hyperuniform distribution determined by a prescribed external potential. The Smoluchowski equation governing the density evolution in 1D is given by
\begin{equation}
\frac{\partial P(r, t)}{\partial t} = \frac{\partial}{\partial r} D \left (\frac{\partial}{\partial r} - \beta f(r) \right ) P(r, t)
\end{equation}
where $f(r) = - dU(r)/dr$. We further consider that the diffusivity $D$ is location independent, which leads to
\begin{equation}
\frac{\partial P(r, t)}{\partial t} = D\frac{\partial^2 P(r, t)}{\partial r^2}  -D \beta \frac{\partial}{\partial r} \left [f(r)P(r, t)\right ]
\label{eq_SM1D}
\end{equation}
The first term on the right hand side of Eq. (\ref{eq_SM1D}) is the standard diffusion term and the second term captures the effects of the external force (field). In the absence of the external field, i.e., $f(r) = 0$, Eq. (\ref{eq_SM1D}) reduces to the normal diffusion equation.

The Fourier transform of Eq. (\ref{eq_SM1D}) is given by
\begin{equation}
\frac{d{\hat P}(k, t)}{dt} = -Dk^2 {\hat P}(k, t) - \beta D (ik) {\sum_{k'}} {\hat P}(k-{k'}, t) {\hat f}({k'})
\label{eq_PK}
\end{equation}
It can be seen from this equation that the dynamics and evolution of the small-$k$ behaviors of ${\hat P}(k, t)$ are coupled to the full spectra of the external force ${\hat f}({k})$. Therefore, it is not immediately obvious how the hyperuniformity of the external field quantitatively determines the dynamics of ${\hat P}(k, t)$ solely based on analytical examination of Eq. ({\ref{eq_PK}}).

In the ensuing discussion, we numerically investigate the evolution of $P(r, t)$ from an initial uniform distribution, i.e., $P(r, t = 0) = \rho_0 = N/L$, and a Gaussian distribution, i.e., $P(r, t = 0) = \rho_0 \exp[-(r-L/2)^2/a^2]$ (with $a = 0.25L$) driven by a stealthy hyperuniform Gaussian potential field at $\beta U_0 = 1$, for which the evolution dynamics is determined by both the diffusion and external forces. These two distinct initial conditions allow us to investigate different dominant dynamics governing the evolution of the density distributions.

The force field resulted from DHU Gaussian potential is given by
\begin{equation}
f(r) = \frac{U_0}{\sigma^2} \sum_{i=1}^M (r-r_i) e^{-\frac{(r-r_i)^2}{2\sigma^2}}
\label{eq_GFF}
\end{equation}
Without loss of generality, we consider a periodic simulation domain with unit length $L = 1$ and all distances are measured with respect to $L$ and made dimensionless. Similarly, the unit of time is chosen such that the dimensionless diffusivity $D = 1$. In this case, the associated Smoluchowski equation is reduced to
\begin{equation}
\frac{\partial P(r, t)}{\partial t} = \frac{\partial^2 P(r, t)}{\partial r^2}  -\sum_{i=1}^M\frac{\partial}{\partial r} \left [\frac{(r-r_i)}{\sigma^2} e^{-\frac{(r-r_i)^2}{2\sigma^2}} P(r, t)\right ]
\label{eq_SM1D2}
\end{equation}
The evolution of $P(r, t)$ given by Eq. (\ref{eq_SM1D2}) is obained by numerically solving the equation via Euler forward method. Specifically, the periodic domain is meshed to $N_{m} = 5000$ grids and the time step is chosen to be $dt = 10^{-8} L^2/D$. The numerical results are verified to be independent of the specific choices of $N_{m}$ and $dt$ values.




We first focus on the uniform initial distribution, i.e., $P(r, t=0) = 1.0$. In this case, the dynamics driving the evolution of density distribution is mainly determined by the forces resulted from Gaussian potential. Fig. \ref{fig_7} shows the evolution of $p(r, t) = P(r, t) - \rho(\beta)$ and Fig. \ref{fig_8} shows the evolution of the associated normalized spectral density $\hat{\psi}(k,t)$. It can be seen that the density distribution is driven by the local forces and rapidly evolve to the equilibrium distribution, indicating fast dynamics

On the other hand, for the Gaussian initial distribution $P(r, t =
0) = \rho_0 \exp[-(r-L/2)^2/a^2]$ (with $a = 0.25L$) the initial
long-wavelength density fluctuations can only be relaxed via the
slower diffusive dynamics. Locally the forces resulted from the
tight-binding potential drive the Brownian particles to distribute
themselves into patterns reminiscent of the local patterns in the
equilibrium DHU distribution. As shown in Fig. \ref{fig_9}, these
local patterns are very rapidly developed, which is mainly driven
by the fast dynamics resulted from the tight-binding potential,
while the overall distribution is still modulated by the initial
Gaussian distribution, which slowly relaxes via diffusive
dynamics. Fig. \ref{fig_10} shows the evolution of the associated
normalized spectral density $\hat{\psi}(k,t)$. It can be seen
that, compared to the uniform initial distribution, the
convergence to the DHU state is much slower in this case, which is
dominated by the slow diffusive dynamics.

Finally, we note that although the two numerical examples studied here are based on the DHU Gaussian potential field derived from stealthy hyperuniform distributions, the insights obtained concerning the fast dynamics driven by the external forces and slow dynamics driven by diffusion also hold for other forms of tight-binding potentials with different classes of hyperuniformity. These analyses also suggest that a desirable equilibrium DHU distribution is much easier to achieve starting from an initial distribution without long-wavelength fluctuations.





\section{Conclusions and Discussion}


We have explored equilibrium DHU states of Brownian particles induced by certain tight-binding potentials which possess the property of hyperuniformity themselves. We analytically showed that hyperuniformity of the external potentials is a sufficient to induce hyperuniform density distributions of the Brownian particles in thermal equilibrium. The evolution of an initial distribution, hyperuniform or non-hyperuniform to the final desirable DHU state driven the potential has also been analyzed numerically.

An interesting observation is that thermal motions in these systems tend to enhance hyperuniformity. In particular, in the limit $\beta = 1/kT = 0$, the density distribution approaches a uniform distribution that is independent of the potential field. We note this observation should be interpreted with the understanding that the density fluctuations relevant to hyperuniformity of the system is associated with the imposed potential field, which is on a much larger length scale than the individual particle level.

By mathematical analogy, the Smoluchowski equation can be considered as a special  case form of the more general class of diffusion-reaction equations \cite{frohner, bode1995pattern, hamik2003excitation}, which have been extensively employed to model a wide spectrum of complex physical and chemical systems. We expect that the insights obtained in this work would also be valuable to design hyperuniform reacting systems. For example, it can be expected that a prescribed hyperuniform distribution of sinks (or sources) would also result in a steady-state hyperuniform pattern in a typical diffusion reaction system. It is also interesting to explore the possibility of realizability of hypothetical DHU patterns via inverse optimization \cite{gommes2012density, gommes2012microstructural}. In addition, the insights obtained for the classical particle systems might also be partially generalized to quantum systems. For example, tight-binding potential models are widely used in electronic properties calculations in solid-state systems, where the electrons are localized near the nuclei. It can be expected that a hyperuniform distribution of nuclei would also lead to a hyperuniform state of electrons, as experimentally observed in certain 2D materials \cite{Ge19, Zh20, chen2021stone}.




Finally, we note that active cells migrating in 3D extra-cellular matrix (ECM) can also be considered as a special type of particles (i.e., cells) influenced by an effective external potential (i.e., the landscape determined by the ECM) \cite{nan2019absorbing, zheng2019modeling, zheng2020modeling, kim2020geometric}. The properties of the ECM, including collagen concentration, fiber orientation and mechanical properties can be computationally designed and experimentally controlled \cite{liang2016heterogeneous, jones2014spatial, jiao2012quantitative, nan2018realizations}, which in turn can influence the dynamics of the multi-cell system. This cell-ECM interaction can in principle be mathematically treated using the general framework based on the Smoluchowski equation derived here, in order to control and achieve possible hyperuniform cellular state \cite{Ji14}.


\begin{acknowledgments}
The author is grateful to Yu Zheng for kind help with numerical
calculations of spectral densities, to Dr. Ge Zhang for providing
numerical realizations of 1D DHU point patterns, and to Dr.
Houlong Zhuang and Dr. Duyu Chen for inspiring discussions and
comments, and to Arizona State University for the generous support
during his sabbatical leave.
\end{acknowledgments}


\end{document}